# Variable Petri Nets for Mobility

Zhijun Ding, *Senior Member*, *IEEE*, Ru Yang, Puwen Cui, MengChu Zhou, *Fellow, IEEE*, and Changjun Jiang

*Abstract*—Mobile computing systems, service-based systems and some other systems with mobile interacting components have recently received much attention. However, because of their characteristics such as mobility and disconnection, it is difficult to model and analyze them by using a structure-fixed model. This work proposes a new Petri net model called Variable Petri Net (VPN) for modeling and analyzing these systems. The definition, firing rule, and related analysis technology of VPN are introduced in detail. In a VPN, the possible interaction interfaces are abstracted as a new kind of places called virtual places, and the occurrences of (dis)connections are described by new functions, which makes it appropriate to describe the component collaboration in systems and realize the scalability and pluggability of systems. Moreover, to overcome the shortcoming that markings cannot reflect link capability of a system, VPNs add a constraint function along with a marking to represent a complete system configuration. Several examples are used to demonstrate the newly proposed model and method.

## I. INTRODUCTION

IN the past few decades, with the continuous promotion and development of (mobile) Internet, many emerging technologies (fields), such as Internet of Thing (IoT), Pervasive Computing and Cloud Computing, have been developed [1]-[3]. These fast growing technologies have been widely used in actual systems, and changed the way people live, work and communicate.

Service-based systems, mobile computing systems and some other systems with mobile interacting components [15] have raised a lot of research interests in recent years. They possess such characteristics as mobility and link dynamicity (frequent disconnection) while facing various environmental changes.

**(1) Service-based systems**.

This work is partially supported by National Natural Science Foundation of China under Grant No. 61672381, National Key Research and Development Program of China under Grant No. 2018YFB2100801 and Fundamental Research Funds for the Central Universities under Grant 22120180508. (Corresponding author: Zhijun Ding.)

Z. Ding, R. Yang, P. Cui, and C. Jiang are with the Key Laboratory of Embedded System and Service Computing, Ministry of Education, Tongji University, and also with the Department of Computer Science and Technology, Tongji University, Shanghai, 201804, China (e-mail: zhijun_ding@outlook.com, yangru@tongji.edu.cn, 1631568@tongji.edu.cn, cjjiang@tongji.edu.cn).

M. Zhou is with the Institute of Systems Engineering, Macau University of Science and Technology, Macau 999078, China and also with the Department of Electrical and Computer Engineering, New Jersey Institute of Technology, Newark, NJ 07102 USA (e-mail: zhou@njit.edu).

SOA (Service-Oriented Architecture) is a coarse-grained, loosely coupled system architecture, and can connect different functional units (services) through the interfaces and contracts defined for them [4]. A service is a component conformant with certain behavior specifications and performs some specific tasks. In the system process based on SOA, several services can be invoked (composed) to accomplish a complicated task. The invoking (composition) of services can be uncertain and dynamic because of different requirements, i.e., the interactions among services (or processes and services) can be dynamic and uncertain.

**(2) Mobile computing systems**.

A mobile computing system consists of several distributed and independent components with a dynamic structure. The mobile robots, agents and nodes in an Ad Hoc network are all practical component examples in this kind of systems [5], [6]. In the execution of these systems, components can move and communicate with each other to accomplish an expected task, and the movement of components or environment change can lead to the dynamic interaction between them. That is, components in a system can connect and disconnect frequently and uncertainly.

In the above systems, mobile components can join or leave the systems dynamically, and communicate (collaborate) with different components. Thus how to describe the multiple collaboration among components and confirm the compatibility, scalability and pluggability become the key to the design of these kinds of systems.

Petri nets (PNs), which own clear graphical representation and are equipped with various analysis methods, have been widely used to design physical structures of various systems. In the traditional modeling method based on PN for systems, such as coloured Petri net (CPN) [12] and Nets-with nets formalization [13], all possible interfaces and possible interaction processes among components should be described in a fixed way. When a component joins or leaves a system, a system model should be modified to describe its possible interaction or disconnection with the system. Thus existing modeling methods are not appropriate to realize the scalability and pluggability of the systems.

In Interface-based Programming [42], an interface refers to an abstraction that a component provides itself to the outside world in order to separate external communication methods from internal operations. Thus an interface should be a separation of definition (specification and constraint) and implementation. Similarly, polymorphism in the C++ program separates the interface from the implementation, and a virtual function can be used to realize polymorphism. The above idea has several advantages: firstly, it makes the program have clear specification and easier to use; secondly, it reduces the coupling of modules in the program, and makes the program more easily maintainable, extensible and pluggable.

Therefore, based on the abstract idea, we introduce a virtual function to the interface (place) in PN to describe the abstraction of the possible interfaces, and then propose a new Petri net called a Variable Petri Net (VPN) to describe mobile systems. A virtual interface (also called a virtual place) is used as a specification of interactions, and also the bridge linking the known and unknown worlds. For example, in Fig. 1(a), a component communicates with the other five through five interfaces, and five components may leave, and new ones may join. In a fixed CPN model as shown in Fig. 1(b), when a component join or leave, an interaction process should be added or deleted, and the model structure has changed; while in VPN as shown in Fig. 1(c), some transitions for interactions are folded, and the possible interfaces are abstracted into a virtual place $I$. Thus a component is added or deleted easily without changing the original model. Therefore, the new proposed model in this work has not only a simpler structure, but also excellent compatibility, scalability and pluggability than a CPN.

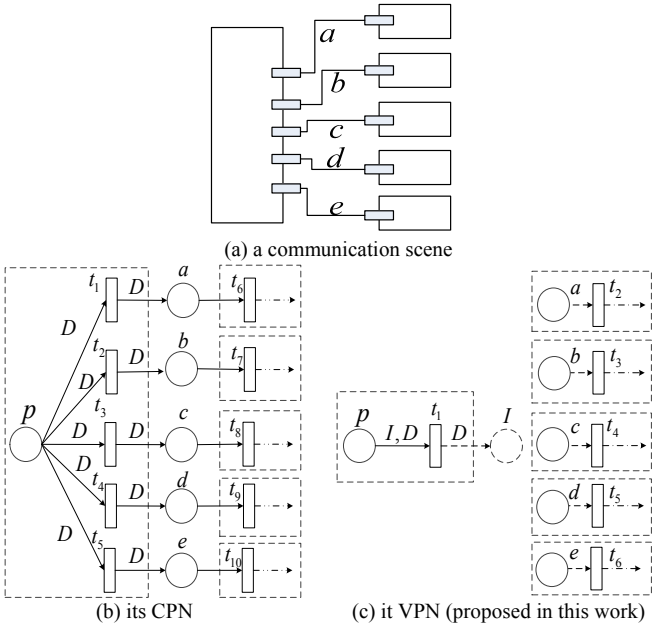

Fig. 1. A scene.

The next section introduces an example to explain VPN. Section 3 gives its definition and firing rule. Section 4 proposes an analysis method and software tool. Section 5 uses two examples to demonstrate them. Section 6 discusses the related work and compares VPNs with other PNs.

## II. MOTIVATING EXAMPLE

Let us consider a common mobile scenario in the daily life. In this scene, there exist several mobile devices. Devices have no physical connection at first and can be connected by different modes, such as Bluetooth and WiFi. Now Device A activates the Bluetooth interface and wants to send a file ($f_1$) through its interface ($I_{A-B}$). Then Device B opens the Bluetooth and discovers the interface $I_{A-B}$ of Device A according to the Bluetooth SDP (Service Discovery Protocol). Then the connection between A and B can be constructed using the interface, and B can receive the file from A as shown in Fig. 2.

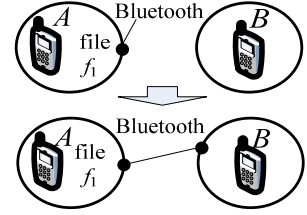

Fig. 2. A practical scene (Example 1).

It is noted that a link (connection) between Devices A and B does not exist at first and is created when they are close and ready to communicate. Hence, as mentioned in the first section, we can think of channels (interfaces) being a virtual place in PNs, which can be instantiated by actual channels (places) when an interaction takes place, and disassociated with those channels when a disconnection takes place. In this example, we use $I$ to denote the virtual channel of the communication.

Example 1 can be modeled as $N_{e1}$ in Fig. 3(a). At first Device B has discovered the interface information of Device A and A wants to send a file, and thus the initial state (marking) of $N_{e1}$ is

$$In\{I_{A-B}\}, St_1\{f_1\}$$

where $In$ is the name of a place for the interface information storage of discovered devices in B and $St_1$ is the name of a place for the file storage in A. $I_{A-B}$ and $f_1$ are names of two tokens in $In$ and $St_1$ which represent the interface information between A and B and a file, respectively.

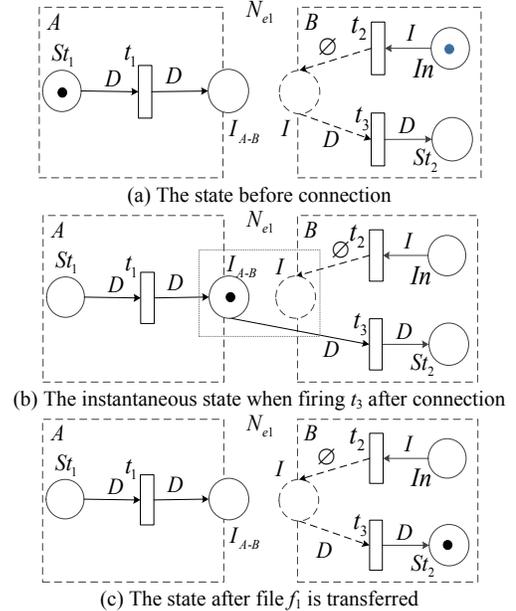

Fig. 3. A new model $N_{e1}$ for Example 1.

Then by firing $t_2$ with the interface information $I_{A-B}$ in place $In$, the connection between B and A can be constructed with mapping (instantiating) of virtual place $I$ to place $I_{A-B}$ ($I\rightarrow I_{A-B}$), and a constraint between $I$ and $I_{A-B}$ is created by using a link function. That is, $I_{A-B}$ associated with $I$ is used as a channel (place) connecting A and B, and a path $St_1$-$t_1$-$I_{A-B}$-$t_3$-$St_2$ can be established. Using this connected place $I_{A-B}$, file $f_1$ can be sent from A to B by firing $t_1$ and $t_3$, and $N_{e1}$ reaches a final state

$$St_2\{f_1\}$$

where $St_2$ is the name of a place for the file storage in B, and $f_1$ is the name of a token for the file in it. It is shown in Fig. 3(c).

Fig. 3 gives a clear description for the connection process in Example 1. In this figure, $I_{A-B}$ is used for both token name and place name. When the virtual place is instantiated to a real one at a transition's firing, the virtual arc between the virtual place and the transition becomes a real (solid) one momentarily and then becomes virtual again after tokens are transferred by the transition and a new state is reached. For example, when firing $t_3$, a solid arc $(I_{A-B}, t_3)$ is generated as shown in Fig. 3(b), and after $f_1$ is transferred to place $St_2$, the arc becomes virtual again.

This new kind of PNs is called a VPN, which we propose in this paper to describe the interfaces, connections and message transfer in mobile systems.

### III. STRUCTURE AND BEHAVIOR OF VARIABLE PETRI NETS

A VPN is a Petri net with a variable structure. It introduces virtual places, makes full use of variable names, and can describe the connection and disconnection in physical systems.

Let $\Sigma$ be a finite set of names which are used to indicate places, tokens and arc weights in the following. Let $\Sigma = V \cup C$, where $V$ is a set of variables, or called formal parameters, and $C$ is a set of constants, or called actual parameters. $C$ contains a special character $\varepsilon$ for the ordinary token by default, and each element $c \in C$ is asscoiated with an arity $n \in \mathbb{N}^+$. For any set $A$, we use $2^A$ to denote all subsets of $A$, $A^*$ to denote all tuples formed by the elements in $A$, and $A^n$ to denote all tuples with exactly $n$ elements of $A$ (whose length is $n$). For example, suppose that $A = \{a_1, a_2\}$, then $2^A = \{\emptyset, \{a_1\}, \{a_2\}, \{a_1, a_2\}\}$, $A^* = \{(a_1), (a_2), (a_1, a_2), (a_2, a_1), (a_1, a_2, a_1),\ldots\}$, $A^1 = \{(a_1), (a_2)\}$.

In the following, we use $\mathbb{N}$, $\mathbb{N}^+$ and $\mathbb{Z}$ to denote the sets of non-negative integers, positive integers and integers, respectively.

**Definition 2.1 (multiset)**. For any set $A$, a **multiset (bag)** $m$ over $A$ is defined as a mapping $m: A \to \mathbb{N}$. The set of all bags over $A$ is denoted by $\mathbb{N}^A$. We use + and - for the sum and difference of two bags, $|m|$ for the number of all elements in $m$ taking into account the multiplicity, and =, <, >, ≤, ≥ for comparisons of bags, which are defined in the standard way.

For example, suppose that $X = \{a, b\}$, then a multiset $m_1$ over $X$ can be defined as $m_1 = \{a, a, b\}$, i.e. $m_1(a) = 2$, $m_1(b) = 1$.

Here we introduce the definition of VPN, which has a changed structure.

**Definition 2.2 (VPN).** A Variable Petri Net (VPN) $N$ over the universe $\Sigma = C \cup V$ is an 8-tuple $N = (P, T, F, \gamma, W, \varphi, \rho, M_0)$, where

1. $P \subseteq C$ is a set of places. Each place is associated with an arity, which is the length of tuples of tokens in it.

2. $T$ is a finite set of transitions. $P \cap T = \emptyset$.

3. $F \subseteq (P \times T) \cup (T \times P) \cup (T \times V) \cup (V \times T)$ is a set of arcs. Each arc $(t, v) \in (T \times V)$ or $(v, t) \in (V \times T)$ is called a virtual arc, and each variable in a virtual arc is called a virtual place.

4. $\gamma: V \to 2^C$ is a constraint function mapping variable $v \in V$ to a set of constants $X \in 2^C$, and each element $c \in X$ is a place in $P$ or a newly generated place in $C$.

5. $W: F \to \mathbb{N}^{(\Sigma^*)}$ is an arc labelling function (expressions) representing the weight for arcs. Each label can be tuples of constants and variables or the empty set $\emptyset$. For any transition $t \in T$, if a variable $v \in V$ meets the condition "$(t, v) \in F$ or $\exists e \in \Sigma: v \in W(t, e)$", it must also satisfy that "$(v, t) \in F$ or $\exists e' \in \Sigma: v \in W(e', t)$". For each arc $f \in (T \times P)$ or $(P \times T)$, $W(f) \in \mathbb{N}^{(\Sigma^n)}$, where $n$ is the arity of the place in $f$.

6. $\varphi: T \to \mathbb{B}$ is a guard function associated with each transition, where $\mathbb{B}$ is the set of all Boolean expressions that can be constructed by using constants and variables in $\Sigma$.

7. $\rho: T \to (\mathbb{B} \times \Theta)$ is a link function of transitions, where $\Theta$ is a series of operations which are done when $\mathbb{B}$ is judged as **true**. For a transition $t$, $\rho(t) = (b, h)$, where $b \in \mathbb{B}$ and $h$ is a do-nothing operation or an operation to add/delete ("+/-") a $\gamma$ constraint to each variable $v$ satisfying that $(t, v) \in F$, denoted as $(v, +/-)$.

8. $M_0$ is an initial marking. A marking of $N$ is a function $M$: $P \to \mathbb{N}^{(C^*)}$, where $M(p) \subseteq \mathbb{N}^{(C^{n_p})}$ ($n_p$ is the arity of $p$) is the set of tokens residing in $p$.

**Remark 2.1.** For a node $x \in P \cup T$, its preset •$x$ and postset $x$• are subsets of $P \cup T \cup V$ such that •$x = \{y|(y, x) \in F\}$ and $x\bullet = \{y|(x, y) \in F\}$. Let $t$ be a transition. For any $x \in \bullet t$ (or $t\bullet$), if $x \in C$, arc $(x, t)$ is called an input arc (or $(t, x)$ is called an output arc); otherwise, $(x, t)$ is called a virtual input arc (or $(t, x)$ is called a virtual output arc). These arcs are collectively called adjacent arcs of $t$. And $x$ is called the (virtual) pre-place or post-place.

**Definition 2.3.** A guard $\varphi(t)$ of a transition $t$ is a relational expression formed by the constants in $C$ and the variables that exist in the input arc expressions or as the virtual pre-places of $t$.

Each token is a tuple of symbolic constants called tuple token or an ordinary black token in conventional PN, which can describe the messages explicitly and directly. The length of a tuple token is equal to or greater than 1 (the parentheses can be omitted if the length is 1) and the ordinary black token $\varepsilon$ is a special case of the tuple token. In particular, if there exists an ordinary token in $p$, the token can be denoted as $p(\varepsilon)$ or $p(\cdot)$. $n$ black tokens in a same place $p$ can be denoted as $p\{\varepsilon,\ldots,\varepsilon\}$ or $p\{n\varepsilon\}$. Each marking $M$ is function mapping each place to a multiset of tuples of constants in $C$. For example, suppose that $C = \{\varepsilon, a, b\}$. Then a token can be $\varepsilon$, $a$, $(a, b)$, etc. Suppose that a place $p$ have two tuple tokens $(a, b)$ at a marking $M$, then $M(p) = \{(a, b), (a, b)\}$, which is a multiset.

All variables in a VPN can be instantiated. That is, if a variable is a virtual post-place or exists in an output arc expression of a transition, it must be a virtual pre-place or exists in an input arc expression of the transition that can be instantiated by the tokens or the constraints when firing the transition. This is a basic rule of VPN, and if this condition is not satisfied, the net cannot be executed. There are two types of arc expressions in a VPN: the tuple of names (variables or constants) and empty set $\emptyset$. The former often denotes the type of tokens (including black tokens) consumed or produced while $\emptyset$ always exists on the output virtual arc $(t, v)$ of transition $t$ and reflects that the transition can bind the variable $v$ with a constant but produces no token.

**Remark 2.2.** In this paper, for the sake of simplicity, if the related guard $\varphi(t)$ of a transition $t$ is constant true, it can be omitted and not given in the definition; if the constraint $\gamma(v)$ of a variable $v$ is $\varnothing$, it can be omitted; if the operation in the link $\rho(t)$ of a transition $t$ is a do-nothing operation, it can also be omitted. If all the mapping relations in a function are omitted, the function can be written as *NULL*. In addition, the special character $\varepsilon$ can be omitted in a set of constants.

**A virtual arc** $(t, v)$ or $(v, t)$ between a variable (virtual place) and a transition means that there exists no actual relation between them initially but possible real one between the transition and a place (constant) as instantiated from the variable in the evolution of a net. The mapping between variables and constants is the key point of our new definition.

Functions $\gamma$ and $\rho$ are two newly proposed ones to give some rules for the mapping between variables including virtual place and constants. $\gamma$ is a constraint function which is used to bind some virtual places with actual places, and can also be regarded as a constraint for the mapping between some variables and constants. $\rho$ is used to modify function $\gamma$. It can add/delete some constraints (bindings) to/from $\gamma$, thus changing the relations between some variables and constants. The guard function $\varphi$ is used to restrict the type (name) of the tranferred data and the interface.

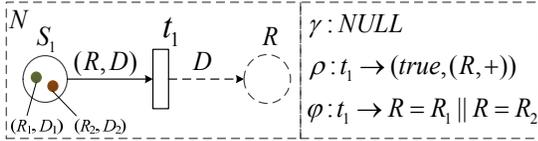

Fig. 4. A simple example.

A VPN can be regarded to have two parts: basic net part and dynamic part. The former means its initial structure, while the latter contains its markings, places and constraints.

Here we use a simple example to explain some elements in the definition of VPN. A VPN $N$ is shown in Fig. 4. In $N$, $C = \{S_1, R_1, R_2, D_1, D_2\}$, $V = \{R, D\}$, and $P = \{S_1\} \subseteq C$. Its basic part is the initial structure shown in Fig. 4. It can be noted that only $S_1$ is used for the real place name and $R$ is used for a virtual place name at first. The guard of $t_1$ means that the firing of $t_1$ can only transfer data to place $R_1$ or $R_2$. If we execute $t_1$ twice, $N$ can have three parts of changes: $R$ is instantiated by two tokens $R_1$ and $R_2$ and new places $R_1$ and $R_2$ are generated (by noting the same name used for tokens and places); the mapping relations between $R$ and $\{R_1, R_2\}$ are added to the constraint function $\gamma$; places $R_1$ and $R_2$ obtain tokens $D_1$ and $D_2$, respectively. Thus the set of places, constraint function and markings are dynamic parts and can change during net execution.

It is mentioned that components in a system may disconnect because of a changing external environment. Example 1 can be extended to Example 2 by considering disconnections. In Example 2 as shown in Fig. 5(a), Device A wants to send two files $\{f_1, f_2\}$ through Bluetooth interface $I_{A-B}$, and Device B can close Bluetooth, disconnect with A and thus suspend the file transfer anytime. The modeling for such disconnection can be realized by using $\gamma$ and $\rho$ in the VPN definition.

The VPN model for Example 2, as shown in Fig. 5(b), is $N_{e2} = (P, T, F, \gamma, W, \varphi, \rho, M_0)$ over $\Sigma = V \cup C$, where $C = \{St_1, St_2, In, De, I_{A-B}, f_1, f_2\}$, $V = \{I, D\}$; $P = \{St_1, St_2, In, De, I_{A-B}\}$, $T = \{t_1, t_2, t_3, t_4\}$, $F = \{(St_1, t_1), (t_1, I_{A-B}), (In, t_2), (t_2, I), (t_2, De), (I, t_3), (t_3, St_2), (De, t_4), (t_4, I)\}$, $\gamma = NULL$, $W = \{(St_1, t_1) \rightarrow D, (t_1, I_{A-B}) \rightarrow D, (In, t_2) \rightarrow I, (t_2, I) \rightarrow \varnothing, (t_2, De) \rightarrow I, (I, t_3) \rightarrow D, (t_3, St_2) \rightarrow D, (De, t_4) \rightarrow I, (t_4, I) \rightarrow \varnothing\}$, $\varphi = NULL$, $\rho = \{t_2 \rightarrow (\mathbf{true}, (I, +)), t_4 \rightarrow (\mathbf{true}, (I, -))\}$, $M_0 = \{St_1\{f_1, f_2\}, In\{I_{A-B}\}\}$. $De$ is a storage place for the interface information of the devices possible to be disconnected from Device B. $t_2$ is to map virtual place $I$ to interface place $I_{A-B}$ for the connection between A and B, and also store the interface information $I_{A-B}$ of A that is possible to be disconnected in $De$. $t_4$ with a $\rho$ function is to unbind $I$ with the disconnected interface place $I_{A-B}$ for the disconnection between A and B. Place $I_{A-B}$ is an available channel for B after firing $t_2$ and becomes unavailable after firing $t_4$.

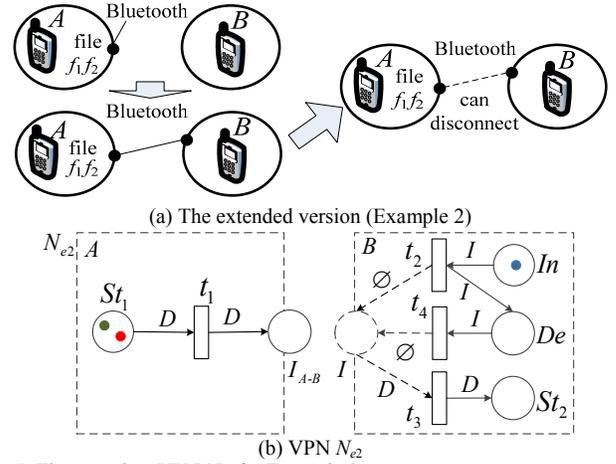

Fig. 5. The complete VPN $N_{e2}$ for Example 2.

Next, given $N = (P, T, F, \gamma, W, \varphi, \rho, M_0)$ over $\Sigma = C \cup V$ being a VPN, some definitions about its execution are introduced.

Variables in the VPN can be substituted with constants during its execution. And these substitutions are defined as bindings as follows.

**Definition 2.4 (Binding).** A binding $\beta$ of any transition $t \in T$ is a (partial) function: $V \rightarrow C$ which associates a variable to a constant, and satisfies that if $v \in V$ is a virtual place such that $(t, v)$ or $(v, t) \in F$, the length of tuples in $W(t, v)$ or $W(v, t)$ should be equal to the arity of place $v[\beta]$, or $W(t, v) = \varnothing$.

A binding exists at one firing of a transition $t$, and it assigns some constants to all variables occurring in virtual pre/post-place names, adjacent arc labels and guard function of $t$. Binding $\beta$ to virtual pre/post-place $v$, adjacent arc labels and guard function $W(p, t)$, $W(t, p)$, $W(v, t)$, $W(t, v)$ and $\varphi(t)$ of $t$ are denoted by $v[\beta]$, $W(p, t)[\beta]$, $W(t, p)[\beta]$, $W(v[\beta], t)[\beta]$, $W(t, v[\beta])[\beta])$ and $\varphi(t)[\beta]$, respectively.

The firing of a transition in a VPN can lead to three types of changes: marking, place set (the set of places) and function $\gamma$. Hence, using a marking, a place set and $\gamma$, we can uniquely identify a configuration of a changed VPN.

**Definition 2.5 (Configuration of VPN).** Let $M$, $P'$ and $\gamma'$ be a marking, a place set and a constraint function of $N$, then $\Pi = (M, P', \gamma')$ is called a **configuration** $\Pi$ of $N$. The initial configuration of $N$ is $\Pi_0 = (M_0, P, \gamma)$.

For example, the initial configuration of $N_{e2}$ is $\Pi_0 = (\{In\{I_{A-B}\}, St_1\{f_1, f_2\}\}, \{St_1, St_2, In, De, I_{A-B}\}, NULL)$.

**Remark 2.3.** If the place set has no change in the execution of a VPN, it can be omitted in each configuration in this paper for simplicity.

Then the firing rule of VPN is introduced as follows.

**Definition 2.6.** A transition $t \in T$ is enabled in a configuration $\Pi = (M, P, \gamma)$ of $N$ iff there exist one binding $\beta$ satisfying that,

1. $\varphi(t)[\beta] =$ **true**;
2. $\forall p \in P, \forall v \in V: (p, t) \in F \Rightarrow M(p) \geq W(p, t)[\beta]$, $(v, t) \in F \Rightarrow (M(v[\beta]) \geq W(v[\beta], t)[\beta]$ and $v[\beta] \in \gamma(v))$;
3. For all variables $v_1, \ldots, v_k$ such that $(v_1, t), \ldots, (v_k, t) \in F$ and $v_1[\beta] = \ldots = v_k[\beta] = p$, $M(p) \geq W(v_1[\beta], t)[\beta] + \ldots + W(v_k[\beta], t)[\beta]$;

which is denoted as $\Pi[t>_\beta$ or $(M, P, \gamma)[t>_\beta$.

In Definition 2.6, if a transition $t$ is enabled with a binding $\beta$, and a virtual arc is involved such that $(v, t) \in F$, there must be a constraint between $v$ and the real place $v[\beta]$ ($v[\beta] \in \gamma(v)$). If a transition $t$ is enabled with a binding $\beta$, all variables in $\varphi(t)$ can be instantiated by $\beta$ and whether the guard is **true** can be determined.

For example, in Fig. 4, $t_1$ is enabled because there exists a binding $\beta_1 = \{R \rightarrow R_1\}$ such that $\varphi(t_1)[\beta] = (R_1 = R_1 \parallel R_1 = R_2) =$ **true** and $M(S_1) > W(S_1, t_1)[\beta]$.

**Definition 2.7.** Firing an enabled transition $t$ with binding $\beta$ at a configuration $\Pi = (M, P, \gamma)$ results in the following changes:

1. $P$ into $P'$: for each constant $v[\beta]$ such that $(t, v) \in F$ and $v[\beta] \notin P$, it is added to the place set $P$ ($P = P \cup v[\beta]$), and $M(v[\beta]) = \varnothing$; The final result of $P$ is denoted as $P'$.
2. $\gamma$ into $\gamma'$: for each variable $v$ such that $(t, v) \in F$, if its condition and operation in $\rho(t)$ is $(b, (v, o))$, **then** $\gamma(v) = \gamma(v) \cup \{v[\beta]\}$ if $b[\beta]$ is **true** and $o =$ "+", and $\gamma(v) = \gamma(v) - \{v[\beta]\}$ if $b[\beta]$ is **true** and $o =$ "–"; The final result of $\gamma$ is denoted as $\gamma'$.
3. $M$ into $M'$ such that for each $p \in P$: $M'(p) = M(p) - W(p, t)[\beta] + W(t, p)[\beta] - \sum_{(v,t) \in F: v[\beta]=p} W(v,t)[\beta] + \sum_{(t,v) \in F: v[\beta]=p} W(t,v)[\beta]$.
4. each arc $(v, t)$ or $(t, v) \in F$ in $N$ into solid arc $(v[\beta], t)$ or $(t, v[\beta])$ at the firing of $t$, and then into virtual arc again when a new marking $M'$ is generated.

The new marking $M'$, place set $P'$, and constraint function $\gamma'$ form a new configuration $\Pi' = (M', P', \gamma')$. Thus, $\Pi$ is transformed to $\Pi'$ by firing $t$ with binding $\beta$, represented by $\Pi[t>_\beta \Pi'$ or $M[t>_\beta M', P[t>_\beta P'$ and $\gamma[t>_\beta \gamma'$.

The first two points of Definition 2.7 describe the change of the place set and the constraint function when firing a transition. It may be slightly obscure, and here we give an explanation. In the first point, if a variable (virtual place) $v$ is instantiated to a constant that does not exist in the place set, this constant is added to the place set. The marking for this constant is initialized as $\varnothing$. This condition describes the creation (or called the discovery) of a new place and binding and is similar to the channel discovery in actual systems. In the second point, if a Boolean expression in $\rho$ of a transition is **true**, the related add/delete operation is executed to modify $\gamma$ by using the binding between each virtual output place of the transition and its instantiation. If "+", the binding is added; if "–", it is deleted. This condition describes the addition and deletion of bindings, and corresponds to connection and disconnection process in some actual systems.

The solid arc instantiated from a virtual arc only exists at the firing of a transition and cannot be added to the net. Here we use the example in Fig. 4 to explain it. If $t_1$ fires twice, $\gamma$ becomes $\{R \rightarrow \{R_1, R_2\}\}$ and $R$ can be instantiated to new places $R_1$ and $R_2$, respectively. This means that two channels are generated one after another. However, arcs $(t_1, R_1)$ and $(t_1, R_2)$ cannot be added to the arc set. If they are added, $t_1$ has two post-places and can produce tokens to them at the same time. This is quite opposite to what we want to express. VPN intends to fold two uncorrelated events (transitions) "sending $D_1$ to channel $R_1$" and "sending $D_2$ to channel $R_2$" to be a transition with a virtual output arc and a virtual post-place. Thus arc $(t_1, R_1)$ or $(t_1, R_2)$, which is used to transfer the data to the place instantiated from $R$, only exists at one firing of $t_1$, and cannot be added to the arc set. That is, new place $R_1$ or $R_2$ has the potential relation with $t_1$, and this relation becomes a real one only when $R$ is instantiated to $R_1$ or $R_2$ at the firing of $t_1$.

**Definition 2.8.** A sequence of transitions $\sigma = t_1 t_2 \ldots t_k$ is a firing sequence if there exists a series of bindings $\beta = \beta_1 \beta_2 \ldots \beta_k$ and configurations such that $\Pi[t_1>_{\beta_1} \Pi_1[t_2>_{\beta_2} \ldots \Pi_{k-1}[t_k>_{\beta_k} \Pi_k$, written as $\Pi[\sigma>_\beta \Pi_k$, and $\Pi_k$ is said to be reachable from $\Pi$ by firing $\sigma$. $\sigma$ can be called a transition (firing) sequence from $\Pi$ to $\Pi_k$; If $\Pi$ is a reachable configuration from $\Pi_0$, then $R(\Pi)$ is the reachability set of all configurations reachable from $\Pi$ ($\Pi \in R(\Pi)$).

**Remark 2.4.** The place in a VPN only can be generated from an element in the set of constants $C$, and $C$ is finite. Thus the place set is a finite set. Similarly, the set of possible bindings when firing a transition is finite.

First we present a firing sequence $t_1$-$t_1$ of VPN in Fig. 4. Its initial configuration is $\Pi_0 = (S_1\{(R_1, D_1), (R_2, D_2)\}, \{S_1\}, NULL)$. Then $t_1$ fires with $\beta_1 = \{R \rightarrow R_1\}$, a new place $R_1$ is added, a new constraint $R \rightarrow R_1$ is created and a new marking is generated, thereby resulting in a new configuration $\Pi_1 = (\{S_1\{(R_2, D_2)\}, R_1\{D_1\}\}, \{S_1, R_1\}, \{R \rightarrow R_1\})$. Then $t_1$ fires again with $\beta_2 = \{R \rightarrow R_2\}$. A new configuration is generated, denoted as $\Pi_2 = (\{R_1\{D_1\}, R_2\{D_2\}\}, \{S_1, R_1, R_2\}, \{R \rightarrow \{R_1, R_2\}\})$.

Then we use $N_{e2}$ in Fig. 5(b) to explain the firing rule. The firing process of a possible firing sequence $t_2$-$t_1$-$t_3$-$t_4$ is presented.

At first, $t_2$ fires with $\beta_1 = \{I \rightarrow I_{A-B}\}$, which results in a new configuration, i.e., $\Pi_0 [t_2>_{\beta_1} \Pi_1$, where $\Pi_1 = (M_1, P, \gamma_1)$, $M_1 = \{De\{I_{A-B}\}, St_1\{f_1, f_2\}\}$ and $\gamma_1 = \{I \rightarrow I_{A-B}\}$ (The connection between A and B is created by $I_{A-B}$).

Then $t_1$ fires with $\beta_2 = \{D \rightarrow f_1\}$, or $\Pi_1 [t_1>_{\beta_2} \Pi_2$, where $\Pi_2 = (M_2, P, \gamma_2)$, $M_2 = \{De\{I_{A-B}\}, St_1\{f_2\}, I_{A-B}\{f_1\}\}$ and $\gamma_2 = \gamma_1$ (file $f_1$ is sent to interface $I_{A-B}$).

Then $t_3$ fires with $\beta_3 = \{I \rightarrow I_{A-B}, D \rightarrow f_1\}$, i.e., $\Pi_2 [t_3>_{\beta_3} \Pi_3$, where $\Pi_3 = (M_3, P, \gamma_3)$, $M_3 = \{De\{I_{A-B}\}, St_1\{f_2\}, St_2\{f_1\}\}$ and $\gamma_3 = \gamma_1$ (file $f_1$ is transferred from A to B).

Finally $t_4$ fires with $\beta_4 = \{I \rightarrow I_{A\text{-}B}\}$, and we have $\Pi_3 [t_4 >_{\beta_4} \Pi_4$, where $\Pi_4 = (M_4, P, \gamma_4)$, $M_4 = \{St_1\{f_2\}, St_2\{f_1\}\}$ and $\gamma_4 = NULL$ (The disconnection between A and B happens). Then in this net, $t_3$ is not enabled any more, and file $f_2$ of A cannot be received by B because of the disconnection. Hence, the dynamic interactions are well depicted.

## IV. ANALYSIS TECHNOLOGY

The incorrectness and unsafety of systems with mobile interacting components may increase the possibilities of system faults and attacks. Hence, the verification of dynamic disconnection and other properties of systems based on VPN is necessary.

### A. Properties of VPNs

VPN is an extended PN, and thus owns similar properties as ordinary PNs do. Hence, based on PN theory, we introduce several basic properties of VPN.

**Definition 3.1 (Place boundness and safety).** Given a VPN $N = (P, T, F, \gamma, W, \varphi, \rho, M_0)$ over $\Sigma$, and any place $p$ in $N$, if there exists $B \in \mathbb{N}^+$ such that $\forall \Pi = (M, P', \gamma') \in R(M_0, P, \gamma)$ and one of the following conditions is satisfied:

(1) $p \in P'$ and $\sum_{i=1}^{s} M(p)(\alpha_i) \leq B$, where $p$ contains $s$ different tokens $\alpha_1, \ldots, \alpha_s$ at marking $M$;

(2) $p \notin P'$;

then $p$ is bounded. The least $B$ satisfying the condition is called the bound of $p$, denoted by $B(p)$:

$B(p) = \min\{B | \forall \Pi = (M, P', \gamma') \in R(M_0, P, \gamma): p \notin P'$ or $\sum_{i=1}^{s} M(p)(\alpha_i) \leq B \}$, where $\alpha_1, \ldots, \alpha_s$ are different tokens in $p$.

If $B(p) = 1$, $p$ is safe. If $p$ is unbounded, $\omega$ is used to denote its bound, denoted by $B(p) = \omega$. Note that $\omega$ is a symbol to represent the infinite number of tokens in places in VPN satisfying "$\omega + a = \omega$, $a \leq \omega$, and $\omega \leq \omega$, $\forall a \in \mathbb{Z}$" [11], [29]-[33].

**Definition 3.2 (VPN boundness and safety).** Let us consider a VPN $N = (P, T, F, \gamma, W, \varphi, \rho, M_0)$ over $\Sigma$, if each place $p$ in $N$ is bounded, $N$ is called a bounded VPN.

$B(N) = \max\{B(p) | \forall \Pi = (M, P', \gamma') \in R(M_0, P, \gamma), p \in P'\}$ is called the bound of $N$. If $B(N) = 1$, $N$ is a safe net.

For example, $N_{e1}$ in Fig. 3 is safe while $N_{e2}$ in Fig. 5(b) is bounded but unsafe.

**Definition 3.3 (VPN deadlock).** Suppose that $N = (P, T, F, \gamma, W, \varphi, \rho, M_0)$ is a VPN over $\Sigma$, and $\Pi_0 = (M_0, P, \gamma)$. $\Pi$ is called a terminal configuration if $\forall t \in T$, $\forall \beta$, $\neg \Pi[t >_\beta$. A VPN is said to have deadlock iff there exists a configuration $\Pi \in R(\Pi_0)$ which is a terminal configuration.

**Definition 3.4 (VPN liveness).** Suppose that $N = (P, T, F, \gamma, W, \varphi, \rho, M_0)$ is a VPN over $\Sigma$, and $\Pi_0 = (M_0, P, \gamma)$. $N$ is live if $\forall t \in T$, $\forall \Pi \in R(\Pi_0)$, $\exists \Pi' \in R(\Pi)$, $\exists \beta$ such that $\Pi'[t >_\beta$.

### B. State analysis

Among various analysis methods of PNs, the state space method is a fundamental and powerful one. From a constructed state space, it is possible to answer a large set of verification questions concerning the state and behavior of a system, such as absence of deadlocks, the possibility of always being able to reach a given state, and the guaranteed delivery of a given service (reachability) [11]-[12].

Hence, to perform the analysis for systems with mobile interacting components based on VPN, we focus on the construction of state space for a VPN to reflect some dynamic behaviors and features of systems. Firstly, we introduce the definition of the reachability tree of a VPN.

**Definition 3.5 (Reachability tree, RT).** The reachability tree $RT$ of a VPN $N = (P, T, F, \gamma, W, \varphi, \rho, M_0)$ over $\Sigma$ is a labeled directed tree whose nodes are the reachable configurations (the root node is $\Pi_0 = (M_0, P, \gamma)$) such that there is an arc from configuration $\Pi$ to configuration $\Pi'$ labeled with $(t, \beta)$, satisfying that $t$ is the fired transition, $\beta$ is the used binding, and $\Pi[t >_\beta \Pi'$.

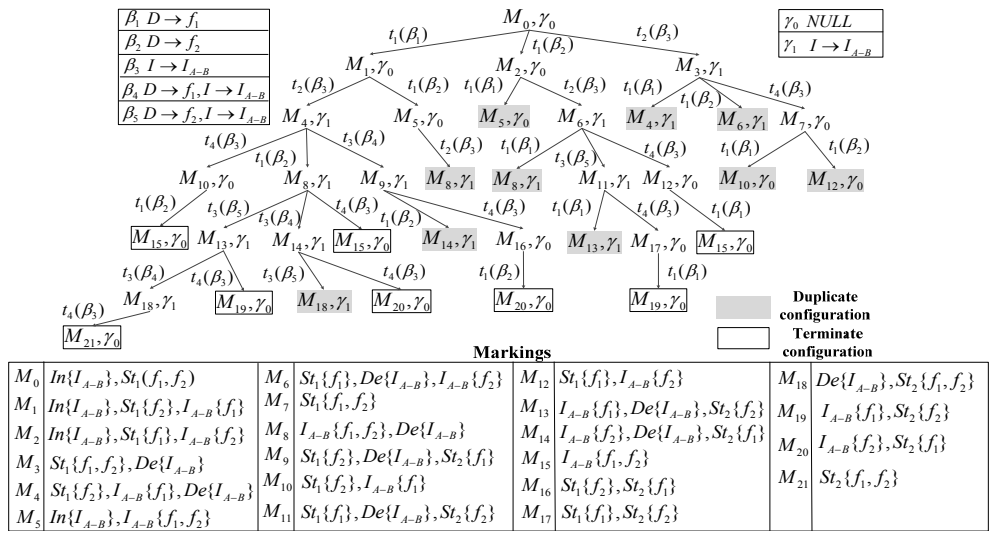

Fig. 6. CT of the VPN $N_{e2}$.

The RT can represent all reachable configurations of a VPN. However, when a VPN is not bounded, its RT becomes infinite. Hence, we use the symbol $\omega$ to represent the infinite number of tokens in places and introduce the configuration tree (CT) of a VPN. The construction algorithm of CT is similar to that of a coverability tree of PNs and we give it in the supplementary file.

CT is a finite directed tree representation of the reachability set, which straightforwardly demonstrates the execution of a net system. By merging two nodes with same configurations, CT can be transformed to a configuration graph (CG). In the CT, there are two kinds of leaf nodes: duplicate and terminal. A duplicate node is a node with a configuration that previously appeared in the tree along the same path. A node is called terminal if no transition can fire at its configuration.

Here we give the example of CT construction. Fig. 6 shows the CT of $N_{e2}$ in Fig. 5(b). In this figure, some subtrees having the same root nodes with another ones are omitted for simplicity. There exist four terminal nodes in Fig. 6, two of which ($M_{19}/M_{20}$, $\gamma_0$) have been mentioned for the interrupted file transfer (only one file was sent successfully) in the previous section, and the last two ($M_{21}$, $\gamma_0$) and ($M_{15}$, $\gamma_0$) mean the successful file transfer (files are all sent successfully) and failed file transfer (files all failed to be sent), respectively.

Because of the information loss resulted from the introduction of $\omega$, the properties of unbounded Petri nets are still difficult to be analyzed based on the CT [31]. Thus we just give three theorems to guarantee some properties of a VPN based on the CT.

**Theorem 3.1 (Finiteness).** The CT of a VPN is finite.

*Proof*: If a CT is infinite, there exists at least one infinite path in it. The infinite path must be an increasing sequence of configurations. In a VPN, the increasing part can be tokens, places or constraints. Because $\Sigma$ is finite, the numbers of places, the types (names) of tokens and the constraints between variables and constants are all finite. Then the infinite part can only be the number of tokens. Consider the most extreme case, and assume that each token $x_i$ can increase infinitely in a path of CT. Then this path always reaches a configuration $\Pi = (M, P, \gamma)$ such that $P$ and $\gamma$ have reached their maximums and for each $p \in P$ and each token $x_i$ in $p$, $M(p)(x_i) = \omega$ (i.e., likely the infinity). According to the construction algorithm of CT, the next reachable configurations of $\Pi$ must be duplicate ones. Thus the assumption is invalid and this path cannot increase infinitely. Other paths can be proved in the similar way. Hence, all paths in the CT are finite, and the CT of a VPN is finite.

**Theorem 3.2 (Reachability).** The CT of a **bounded** VPN consists of only but all reachable configurations from its initial configuration.

*Proof*: The proof of the theorem is obvious from the construction algorithm of CT. Clearly, all configurations in CT are reachable in a bounded VPN and all reachable configurations are contained in CT.

**Theorem 3.3 (Deadlock).** A **bounded** VPN has a deadlock if and only if its CT has a terminal node.

*Proof*: If the CT of a bounded VPN has a terminal node, no transition can fire at the configuration of this node, and this configuration must be reachable in the VPN according to Theorem 3.2. Thus this configuration is a terminal one in the VPN and VPN has a deadlock; If a bounded VPN has a deadlock, there must exist a reachable terminal configuration. According to Theorem 3.2, its CT should contain the node with this configuration and thus have a terminal node.

Therefore, CT can be used for the reachability and some other properties' analysis for VPNs and thus actual systems.

### C. Behavior analysis

Then based on the newly introduced properties and state analysis method, VPN can be used to give the related behavior analysis, such as connection analysis, which may be hard for some existing models, such as CPN [12]. Firstly, some definitions about the behaviors of VPN are proposed. Suppose that $N = (P, T, F, \gamma, W, \varphi, \rho, M_0)$ is a VPN over the universe $\Sigma = C \cup V$, and the initial configuration $\Pi_0 = (M_0, P, \gamma)$ unless otherwise stated in the following discussion.

**Definition 3.5.** $\Gamma(N)$ satisfying that

1) $\gamma \in \Gamma(N)$, and
2) $\Gamma(N) = \{\gamma' | \exists \sigma, \beta: (M_0, P, \gamma)[\sigma>_\beta (M', P', \gamma')\}$,

is called the **connectivity set** of $N$.

$\Gamma(N)$ contains all possible constraint functions of $N$. The possible constraint function sequences from the initial $\gamma$ of $N$ can be generated from its CT (CG) and $\Gamma(N)$.

**Definition 3.6.** The precedent difference between two constraint functions $\gamma'$ and $\gamma''$ is defined as $\gamma''' = \gamma' - \gamma''$, which is a function mapping $V$ to $2^P$ satisfying that for each $v \in V$, $\gamma'''(v) = \gamma'(v) - \gamma''(v)$ that is a set of the constants in $\gamma'(v)$ but not $\gamma''(v)$. Similarly, the subsequent difference between them is defined as $\gamma'' - \gamma'$.

The difference between two constraint functions can denote the new links (bindings) and broken links between variables and constants in the net execution.

**Definition 3.7.** The set of possible newly created links and broken ones in $N$ are defined as **C-set** and **B-set**, and the set of links always maintained or never broken after their creation in $N$ is defined as **A-set**. Each element of them is a mapping from a variable $v \in V$ to one constant $c \in C$ or a group of constants $G = \{c_1, c_2, .. c_l\} \subseteq C$, denoted by "$v \to c$" or "$v \to G$". That is, each set can also be regarded as a function: $V \to 2^C$. **C-set**, **B-set** and **A-set** constitute a link tuple $\mathbb{L} = ($**C-set**, **B-set**, **A-set**$)$.

The link tuple can be obtained based on CG and connectivity set, and we give the algorithm to discover it in the supplementary file. For example, $\mathbb{L}$ of Example 2 in Fig. 5 can be computed according to the algorithm. Its $\mathbb{L} = ($**C-set** $= \{I \to I_{A-B}\}$, **B-set** $= \{I \to I_{A-B}\}$, **A-set** $= \varnothing)$, which reflects that the link between Devices A and B based on the Bluetooth interface $I_{A-B}$ can be created and also disconnected.

Then based on the link tuple, which links are newly created, disconnected or maintained can be answered. This can be used to verify the connectivity (interaction) or the change of the context in systems.

## V. CASE STUDY

In this section, we study two practical systems (a services-based system [23], [24] and a vehicular cyber-physical system [34]) to show the usefulness of VPN and how to analyze it.

### A. A Services-based system (***Program: GymLocker* [23]**)

Business Process Execution Language for Web Services (WS-BPEL) can specify business processes by integrating Web

Services to realize desired functionality. In recent years, it is becoming an industrial standard of Web Services composition, and intensive efforts have been made on related research and implementation. The language has been revised with a new published version, and many WS-BPEL execution engines have been developed, such as BPWS4J [23], [24].

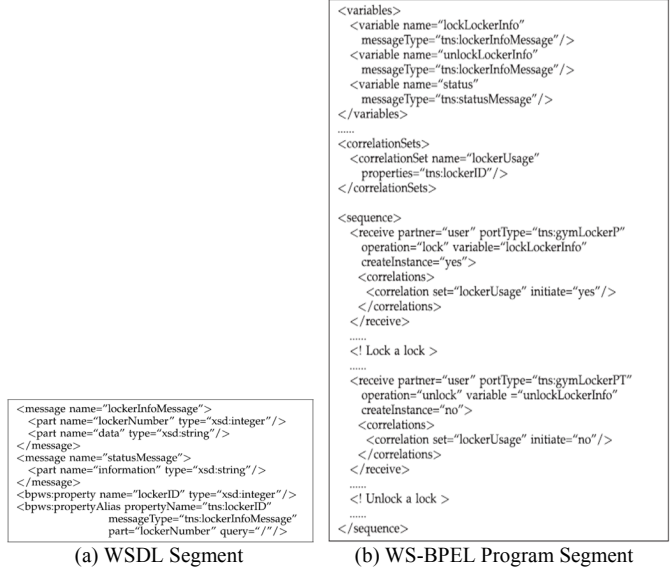

Fig. 7. GymLocker Program (Example 3).
(a) WSDL Segment  (b) WS-BPEL Program Segment

**GymLocker** (shown in Fig. 7(a) and (b)), which is a sample program of BPWS4J defines a process of the management behavior of a lock, including (1) locking up a lock and (2) unlocking a lock. In this program, the process locks up a lock after receiving a *lockLockerInfo* message in the first message reception of Activity **Receive** through the *PortType* (interface), and assigns a constant to a variable *status* by Activity **Assign**, and then return a result by Activity **Reply**. Similarly, the process unlocks a lock after receiving an *unlockLockerInfo* message in the second message reception of Activity **Receive** and then **Reply** (shown in Fig. 7(b)).

A correlation set contain a set of properties that can route the received message to the correct instance. Correlation set *lockerUsage* correlates *lockLockerInfo* and *unlockLockerInfo* by sharing their first part *lockerNumber* (shown in Fig. 7(b)). That is, if a lock is locked up after receiving a *lockLockerInfo* in a BPEL instance, it can only be unlocked after receiving an *unlockLockerInfo* whose *lockerNumber* value is the same as that of the preceding *lockLockerInfo*. The mismatching of *lockerNumber* can result in a fault in a BPEL program.

Here we use VPN to construct a model for the whole program, generate its CT and then verify its properties.

**1. Modeling process**

The whole WS-BPEL program of **GymLocker** is modeled as a VPN model $N_{e3}$ in Fig. 8. It is defined as:

(1) $P$, $T$, $F$, $\gamma$, $W$, $\varphi$ and $\rho$ is obvious in Fig. 8, and $M_0 = \{v1\{(lockInfo, locknumber, data)\}, user\{(client, gymLockerPT)\}, correlationset\{lockerUsage\}, lockLockerInfo\{(lockLockerInfo, null, null)\}, status\{(status, null)\}, unLockLockerInfo \{(unLockLockerInfo, null, null)\}, v2\{ok\}, v4 \{(unLockLockerInfo, lockNumber1, data1)\}, initial(\varepsilon)\}$.

(2) $C = \{initial, user, correlation set, lockLockerInfo, status, unLockLockerInfo, final, s1, s2, s3, s4, s5, v1, v2, v3, v4, v5, c1, c2, c3, c4, lockNumber, lockNumber1, data, data1, ok, client, null, lock, unlock, gymLockerPT, lockerUsage, fault\}$; $V = \{PortType, CName, PName, Information, Information1, LockNumber, LockNumber1, Property, Variable, Data, Data1\}$.

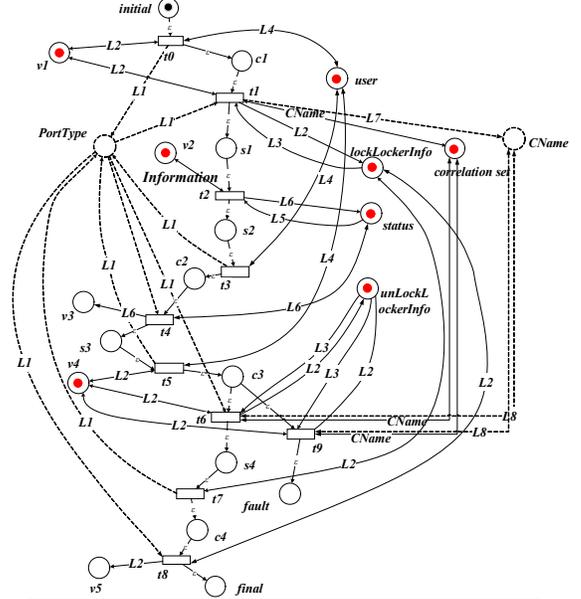

(a) The VPN model

| | | |
|---|---|---|
| Transitions | $t_0, t_1$ | **Receive** Activity: receives some variables and creates the correlationset |
| | $t_2$ | **Assign** Activity: assigns some constants (variables) to variables |
| | $t_3, t_4/t_7, t_8$ | **Reply** Activity : return a result |
| | $t_5, t_6, t_9$ | **Receive** Activity: receives some variables and verifies the correlationset |
| Places | $c1, c3$ | Control places of two **Receive** Activities |
| | $c2, c4$ | Control places of two **Reply** Activities |
| | $s1, s2, s3, s4$ | Connection places between activities of BPEL process |
| | correlation set | Correlation set place of BPEL process |
| | initial,final,fault | State places of BPEL process |
| | user | Place storing PartnerLinks (indicating interacting parties) of BPEL process |
| | $v1, v4$ | Places storing the messages sent to BPEL process |
| | $v2$ | Place storing constants in **Assign** Activity |
| | $v3, v5$ | Places receiving messages from **Reply** Activity |
| Variables (virtual place) | Variable | Particularly indicating the variable (message) called by BPEL activity |
| | PName | Name of some PartnerLink (referred to as Pname ) |
| | PortType | The virtual interface |
| | CName | The correlation set name |

(b) Usage of some places, transitions and variables

Fig. 8. VPN model $N_{e3}$ for the WS-BPEL Program "GymLocker".

The VPN model $N_{e3}$ is constructed based on the WS-BPEL program of GymLocker. The detailed and specific model construction for each activity in the BPEL program can be found in [41], and we just give a simple explanation here. In this model, there exist two virtual places *PortType* and *CName*. *PortType* is the variable for the interface between a BPEL process and clients or external services, and it can transfer its name and message as (*PortType*, *Variable*). *CName* is the variable for the correlation set name which can be initialized or matched in the program execution using the tuple (*CName*, *LockNumber*) or (*CName*, *Property*). *lockLockerInfo* and *unlockLockerInfo* are places for specific variables that own three parts (*Variable*, *LockNumber*, *Data*) or (*Variable*, *LockNumber1*, *Data1*). *status* has one part (*Information*) or (*Information1*).

**2. Analysis process**

To automate VPN modeling and analysis for an actual system, we have developed a computer tool called VPN Tool. This tool is implemented in Java on the MyEclipse platform, and can run in the Windows environment. It has several functions, including VPN drawing, incidence matrix generation, configuration tree generation, and deadlock detection.

Then we use our tool to draw $N_{e3}$ and construct its CT. The total number of different nodes in the CT is 11, and we just give a small part here (Fig. 9(a)). According to the CT of $N_{e3}$, 2 deadlocks are found in our tool (Fig. 9(b)). One deadlock (deadlock1) means the normal termination of the BPEL program, while the other deadlock (deadlock0) means a fault. It has been mentioned that the correlation set confirms that one lock can only be unlocked by an unlock message in the same instance. Thus deadlock0 is detected because of the fault of the correlation set mismatching of two lock and unlock messages in different instances. We can use some deadlock control methods in the system and program design to control such deadlocks [25]-[27].

In addition, the link tuple $\mathbb{L}$ of $N_{e3}$ is (**C-set** = {*PortType* → *gymLockerPT*, *CName* → *lockerUsage*}, **B-set** = ∅, **A-set** = {*PortType* → *gymLockerPT*, *CName* → *lockerUsage*}). The first element in **C-Set** can reflect the interaction between BPEL and clients (or services) (link change), while the second one reflects a new initialized correlation set during BPEL execution (context change).

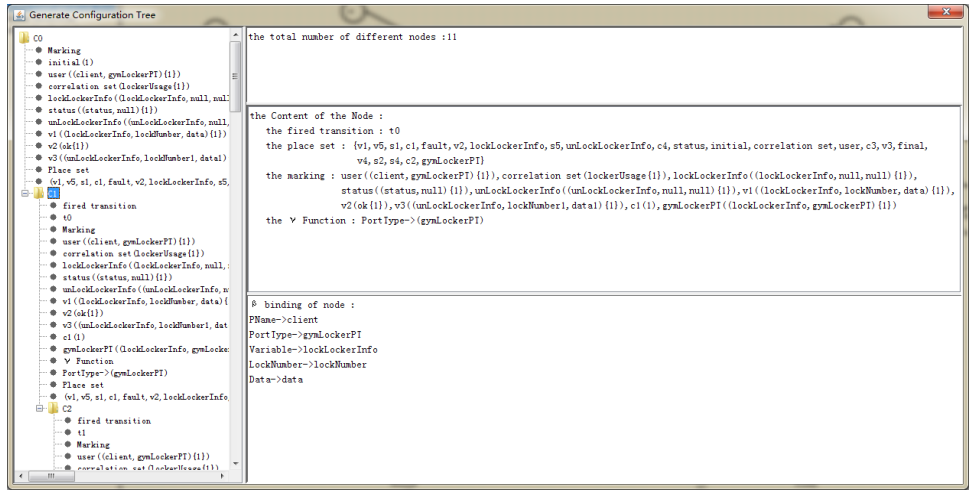

(a) A part of CT for $N_{e3}$ in VPN Tool

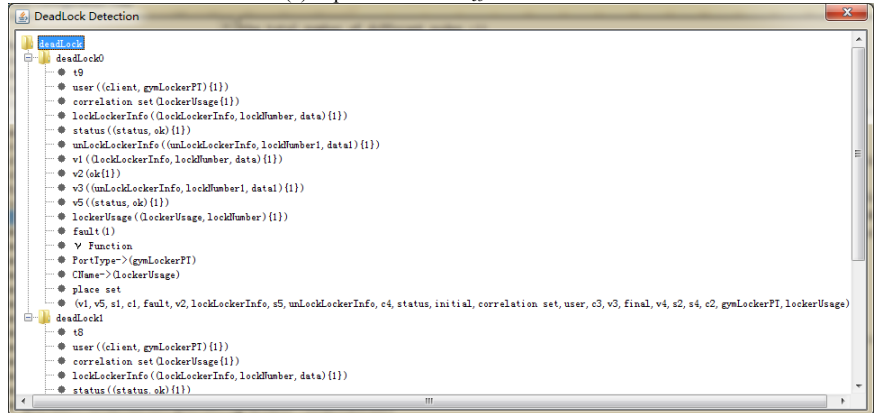

(b) Two deadlocks in VPN Tool

Fig. 9. The modeling and analysis of $N_{e3}$ based on VPN Tool.

### B. A Vehicular Cyber-Physical System (VCPS)

A vehicular Cyber-Physical System (VCPS) is an emerging research field. This kind of systems can provide environmental information for vehicles, and improve public transportation's efficiency through the communications among vehicles through a network system.

In a VCPS, vehicles (cars) can move, cooperate and exchange the environmental information they have gathered. Here we consider an abstracted scenario, in which there exist several groups (clusters). Each group has a control car (leader) and several slave cars (members). On the one hand, the control car can receive a request (REQ) for the environmental data from the internal slave cars (in the same group), and response the environmental data to them (RES); on the other hand, it can perceive the approach of external slave cars (of another group), sends the request to them, and then receive the latest data from them. Slave cars cannot move to another place before receiving the related environmental data from their control car.

For simplicity, supposing that there exist two groups $G_1$ and $G_2$ only. $G_1$ includes a control car $C_A$ and a slave car $S_A$, and $G_2$ includes a control car $C_B$ and a slave car $S_B$. Then this simplified scenario is shown in Fig. 10. It is noted that there exists dynamic interactions between control and (internal or external) slave cars.

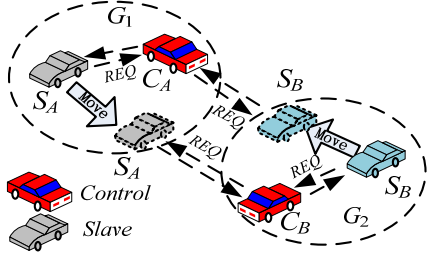

Fig. 10. A Vehicular Cyber-Physical System (Example 4).

VPN specializes in the modeling for dynamicity in a mobile context, and thus is an appropriate model for this system. In the following, we use VPN to model and analyze this system.

**1. Modeling process**

In the modeling, two control cars are modeled as one entity (component) $\Delta$, and two slave cars are modeled as $\Psi$. Then VPN $N_{e4}$ is shown in Fig. 11. It is defined as: $N_{e4} = (P, T, F, \gamma, W, \varphi, \rho, M_0)$ over $\Sigma$, where

(1) $P, T, F, \gamma, W, \varphi$ and $\rho$ is obvious, and $M_0 = \{O_{S1}\{(S_A, D_{A-S}), (S_B, D_{B-S})\}, O_{C1}\{(C_A, D_{A-C}), (C_B, D_{B-C})\}, In\{(S_A, I_A, O_A), (S_B, I_B, O_B)\}, Gr\{(C_A, G_1), (C_B, G_2)\}, Th_1\{(C_A, G_1), (C_B, G_2)\}, Th_2\{S_A, S_B\}\}$.

(2) $\Sigma = C \cup V$ where $C = \{C_A, C_B, S_A, S_B, G_1, G_2, O_{C1}, O_{S1}, D_{A-S}, D_{B-S}, D_{A-C}, D_{B-C}, In, I_A, O_A, I_B, O_B, Gr, Th_1, Th_2, Ne, REQ, RES, HEA, SLA, NUL\}$; $V = \{L, N, I, O, D, N', D', RE, TY\}$.

Fig. 11 also shows the usage of several constants (places) and transitions in $N_{e4}$. The data transferred in the interaction is denoted as an 8-tuple $(RE, N, N', L, I, O, D, TY)$ in the figure, where $RE$ is used for data type, $N$ or $N'$ for the control/slave car, $L$ for the group, $I$ and $O$ for virtual interfaces, $D$ for data and $TY$ for the message sender (source). The transitions $t_1$-$t_8$ (except $t_3$ and $t_5$) can generate new places (interfaces) by the instantiation of virtual places $I$ and $O$ when two cars interact. Thus the dynamicity and mobility of this system can be directly and vividly described.

Moreover, if the scenario in Fig. 10 is extended by adding or deleting groups or cars, model $N_{e4}$ can be easily extended by adding or deleting more tokens in several corresponding places. Thus the VPN model for this system has fine scalability.

**2. Analysis process**

We can use the VPN tool to draw $N_{e4}$ and construct its CT. According to the CT of $N_{e4}$, no deadlock is found. This means that the control cars and slave cars in two groups work coordinately, and thus the VCPS in Fig. 10 has good coordination.

In addition, the link tuple $\mathbb{L}$ of $N_{e4}$ is computed as (**C-set** = $\{I \rightarrow \{I_A, I_B\}, O \rightarrow \{O_A, O_B\}\}$, **B-set** = $\varnothing$, **A-set** = $\{I \rightarrow \{I_A, I_B\}, O \rightarrow \{O_A, O_B\}\}$). This means that the link between the control car ($C_A$ or $C_B$) and slave car ($S_A$ or $S_B$) can be constructed and maintained without disconnection, and thus the VCPS in Fig. 10 has fine connection (communication).

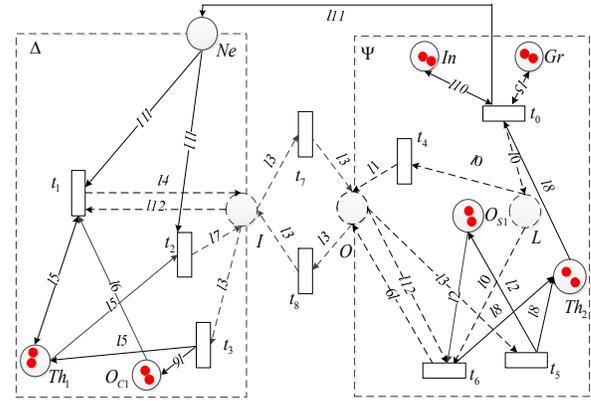

(a) The VPN model

| | Usage |
|---|---|
| REQ/RES | Represent that the message type is data request or response. |
| HEA/SLA | Represent the message sender is control (head) or slave car. |
| $O_{C1}, O_{S1}$ | Store the message for the data RES of (control or slave) cars |
| In | Stores the interfaces for the communication of the slave cars |
| Gr | Stores the groups |
| $Th_1, Th_2$ | Confirm a single execution of (control or slave) cars |
| Ne | Perceives the movement of the slave car |
| $t_0$ | Movement of slave car |
| $t_1$ | Control car receives the data REQ from and sends the data RES to internal slave car. |
| $t_2, t_3$ | Control car sends the data REQ to and receives the data RES from external slave car, and stores the data. |
| $t_4, t_5$ | Slave car sends the data REQ to and receives the data RES from internal control cars, and stores the data. |
| $t_6$ | Slave car receives the data REQ from and sends the data RES to external control cars. |
| $t_7, t_8$ | The interaction (send-receive) mechanism between control and slave cars |

(b) The meaning of some constants, places and transitions in VPN model
Fig. 11. The VPN model $N_{e4}$ for the example in Fig. 10.

Hence, from previous examples, the modeling and analysis process based on VPN can be presented rather directly and clearly. It is noted that VPN can be used for not only the more precise modeling of systems with mobile interacting components, but also the property analysis such as dynamic interaction and context analysis of systems, which has been rarely seen in the existing PN studies.

## VI. RELATED WORK

In recent years, various formal methods have been developed to model and analyze systems with mobile interacting components.

Process algebra methods, such as CCS (Calculus of Communicating Systems) [7], π-calculus [8], the generalizations as Distributed Join calculus [9] and Ambient calculus [10], have gained certain development. They have a solid algebraic foundation and own a rich set of tools, which make them useful in describing the system having a dynamic topology. However, they have no intuitive and hierarchical graph structure, and thus cannot be analyzed using the graphic techniques.

A bigraphical reactive system (BRS) consists of a bigraph and a set of reaction rules as proposed by Milner [35]. It emphasizes both locality and connectivity that can be used to simulate the location and interconnection of mobile system entities, and can be reconfigured by using its reaction rule. It is a powerful model for mobile systems. However, although they have the graphical representations, their foundations are process algebras and logical reasoning.

When modeling a system based on a PN, the system execution and state are both reflected in its graphical structure. Thus we want to study the connections and disconnections in mobile systems from a new perspective of PNs in this work. Compared with BRS, VPN in this work is still a PN, and it has a simpler formation and is easier to operate. The actions and states are involved in it directly and are an interactive whole instead of separate parts. Moreover, VPN introduces several new functions to record and analyze the connections and disconnections, and to discover new channels.

PNs are a classic formal method [11]. They have the graphical representation, can describe physical structures of the modeled systems intuitively, and are equipped with various analysis methods. Based on PNs, much work has been done for the modeling and analysis of systems with dynamicity.

A CPN [12] is an excellent extension of PNs. In a CPN, each token has attached a data value called the token colour, which makes nets possible to use data types and complex data manipulation. Token colours can describe the dynamicity and variability of resources.

Valk proposed the formalization of the "nets within nets" (Elementary object system) approach to dynamic systems [13]. The "nets within nets" formalization is a hierarchical PN in which tokens can also be PNs (token net) instead of black dots. The nesting of token nets for components into other nets can reflect the dynamicity of components. Based on "nets within nets", a Nested Petri net (NPN) has been developed [14], [15]. NPNs can simulate Petri nets with reset arcs [36]. Petri nets with reset arcs extend the basic model with special "reset" arcs, which denote that the firing of some transitions resets (empties) the corresponding places. Some properties such as reachability and boundedness are undecidable for these nets. Xu et al. proposed a two-level approach based on Predicate Transition nets (PrT nets) to model and verify mobile agent systems [16], [17]. A system is divided into several parts, such as an environment part and connectors, and each part is modeled by a PrT net. Agent nets are encapsulated as parts of tokens and can be moved. This approach can model and analyze the dynamicity (mobility) of agents. It is noted that the above ones mainly use the net execution, net hierarchy and new labels to denote system dynamicity (mobility).

Haddad and Poitrenaud developed a model called the recursive Petri nets (RPN) for dynamic structure of processes [37]. An RPN has the same structure as an ordinary PN except that the transitions are partitioned into two categories: elementary and abstract transitions. The reachability and finiteness of RPNs remain decidable. They are enable to model complex mechanisms of DESs, such as interrupts, fault-tolerance and environment-driven behaviors. Yet RPNs have a fixed structure and fail to describe the specific data (messages). Then Kheldoun et al. proposed a model called Recursive ECATNets (RECATNets) combining abstract data types and RPNs [40]. This model can describe a set of Business Process Modeling Notation (BPMN) features such as cancellation and multiple instantiation of subprocesses. Based on the state space of RECATNet, some properties such as reachability and deadlock can be analyzed, which is similar to VPN. Different from RECATNet, VPN focuses on describing dynamic interactions and context by using dynamic structures. It introduces the concepts of configurations that can record states, linking capabilities and new channels of systems, and thus provides more space for the verification of the properties related to interactions.

Some researchers concentrated on the transformation relations between process algebras and Petri nets [18], [38]-[40]. Asperti and Busi proposed the notions of Mobile Petri net (MPN) and Dynamic Petri net (DPN) [18] based on the process algebra. These methods add mobility to PNs, and determine the postset of transitions according to their preset. They offer a direct way to express systems with a changing structure in which components can be dynamically linked to others, possibly depending on previous communications. A Reconfigurable Net, as another important branch of PNs, has gained much development [19]-[22]. This kind of net and its extensions, such as Reconfigurable Object Petri net (ROPN), can dynamically modify their own structure by rewriting some of their components according to rewriting rules. Thus they can model systems that change their structures dynamically. It is noted that they mainly use the changed net to denote the dynamicity of systems.

All of these methods have unique features and can be used in some applications of dynamic systems. However, they have paid inadequate attention to dynamic (dis)connections led from the components' dynamic join and quit, and thus are weak in realizing the scalability and pluggability of systems. In addition, there is a vacancy in the studies of analysis techniques especially behavior analysis for systems with mobile interacting components. A table to show the overall comparison of some PN models has been given in the supplementary file.

We can summarize main differences in both mathematical and physical meanings between VPN and other PN models as three points:

**1) Folding of transitions**

Those high-level PNs, such as CPN and MPN, add a data type to ordinary tokens, and use colored tokens to represent the variability and dynamicity of tokens (resources). Then they can fold places in ordinary PN by distinguishing difference colors

(types) of tokens. Following the similar idea, VPN adds the function mapping to places, and then folds some transitions. At this level, VPN can be regarded as the further folding of CPN, and thus has a simpler structure than CPN as seen in Fig. 1. More explanations for the translation relations among PN, CPN and VPN are given in the supplementary file.

**2) Introduction of virtual place**

In most high-level PN models, places and flow relations between places and transitions are fixed and pre-defined, which makes them difficult to describe the joining or leaving of components directly. VPN introduces a virtual place to abstract all possible interaction interfaces, and also two functions called $\gamma$ and $\rho$ to give some constraints to a virtual place. On one hand, a virtual place can bind/unbind with some known interfaces by using $\gamma$ and $\rho$, which corresponds to the connections/ disconnections among components; on the other hand, it can be used to discover new interfaces led from the joining of some components or other contextual changes. This makes VPN be powerful to describe the changed interactions in systems and confirm the scalability and pluggability of systems.

**3) Configuration.** VPN uses a configuration containing a marking (a vector describing the net tokens), a place set and a function $\gamma$ to describe the net condition, and also can serialize them, while other high-level PNs mostly use the markings only.

Markings only are insufficient to VPN to represent the system situations because they do not reflect the linking capabilities of a system. A configuration of VPN, in which a marking can reflect the system running state and a function $\gamma$ can reflect the linking capabilities of the system, can describe a situation of a system thoroughly. This can help us generate the state sequence (marking sequence) and link sequence (constraint function sequence), which can be used for the further system analysis.

Based on the above characteristics, VPN is more appropriate to model the dynamic links and contextual changes in systems.

## VII.  CONCLUSION

This paper addresses the modeling and analysis issue for systems with mobile interacting components based on PNs. Firstly a new PN model called Variable Petri Net (VPN) is proposed. VPN introduces a function to places to describe interfaces, and new functions $\gamma$ and $\rho$ to describe the occurrences of (dis)connections, and is thus able to handle dynamic links and the join/leave of components in systems, , which is highly difficult, if not impossible, with the existing PN models [43]-[45]. Then the related analysis techniques including the graphic analysis and behavioral analysis of VPN are presented. Several examples are used to explain and demonstrate VPN, and the relations and differences between VPN and other extended PNs are discussed as well. VPN has direct and specific modeling ability than other PNs in the aspect of dynamicity and also scalability, and is appropriate to describe the distributed systems composed of components with changing relations.

In the future work, we intend to focus on the algebraic analysis methods and temporal property of VPN [28].

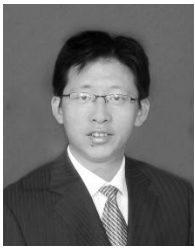
**ZhiJun Ding** received the M.S. degree from Shandong University of Science and Technology, Taian, China, in 2001, and Ph.D. degree from Tongji University, Shanghai, China, in 2007.

Now he is a Professor of the Department of Computer Science and Technology, Tongji University. His research interests are in formal engineering, Petri nets, services computing, and mobile internet. He has published more than 100 papers in domestic and international academic journals and conference proceedings.

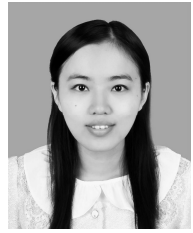
**Ru Yang** received the B.S. degree from Shandong University of Science and Technology, Qingdao, China, in 2013. She is currently pursuing the Ph.D. degree with the Department of Computer Science and Technology, Tongji University, Shanghai, China.

Her current research interests include Petri nets and formal engineering.

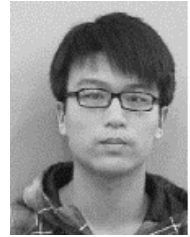
**Puwen Cui** received the B.S. degree from HoHai University of Water Conservancy and Hydropower Engineering, Nanjing, China, in 2016.

He is currently pursuing the M.S. degree with the Department of Computer Science and Technology, Tongji University, Shanghai, China.

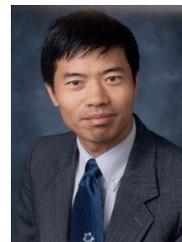
**MengChu Zhou** (Fellow, IEEE) received the B.S. degree in control engineering from the Nanjing University of Science and Technology, Nanjing, China, in 1983, the M.S. degree in automatic control from the Beijing Institute of Technology, Beijing, China, in 1986, and the Ph.D. degree in computer and systems engineering from the Rensselaer Polytechnic Institute, Troy, NY, USA, in 1990. He joined the New Jersey Institute of Technology (NJIT), Newark, NJ, USA, in 1990, where he is currently a Distinguished Professor of electrical and computer engineering. His research interests are in Petri nets, intelligent automation, the Internet of Things, big data, web services, and intelligent transportation. He has over 800 publications including 12 books, 500 journal articles (over 400 in the IEEE TRANSACTIONS), and 29 book-chapters and holds 26 patents. He is a Life Member of Chinese Association for Science and Technology, USA, and served as its President, in 1999. He is a fellow of International Federation of Automatic Control (IFAC) and American Association for the Advancement of Science (AAAS). He was a recipient of the Humboldt Research Award for U.S. Senior Scientists from Humboldt Foundation, the Franklin V. Taylor Memorial Award, and the Norbert Wiener Award from the IEEE Systems, Man, and Cybernetics Society. He is the Founding Editor of the IEEE Press Series on Systems Science and Engineering.

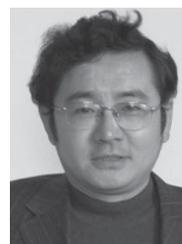
**ChangJun Jiang** received the Ph.D. degree from the Institute of Automation, Chinese Academy of Sciences, Beijing, China, in 1995.

He is currently a Professor with the Department of Computer Science and Technology, Tongji University, Shanghai, China. His current research interests include concurrency theory, Petri nets, formal verification of software, cluster, grid technology, program testing, intelligent transportation systems, and service-oriented computing. He has published more than 100 publications.